\documentclass{article}

\usepackage{arxiv}

\usepackage[utf8]{inputenc} % allow utf-8 input
\usepackage[T1]{fontenc}    % use 8-bit T1 fonts
\usepackage{amsfonts}       % blackboard math symbols
\usepackage{nicefrac}       % compact symbols for 1/2, etc.
\usepackage{microtype}      % microtypography
\usepackage{lipsum}
\usepackage{graphicx,amsmath,amssymb}
\usepackage{geometry}
\usepackage{longtable}
\usepackage{booktabs}
\usepackage{soul}
\usepackage{tabularx}
\usepackage{ltablex}
\usepackage{xcolor}
\usepackage{color}
\usepackage{braket}
\usepackage{footnote}
\usepackage{xr}
\usepackage{prettyref}
\usepackage{url}
\usepackage{float}

\title{Enhancing countries’ fitness with recommender systems on the international trade network}

\author{
  Hao Liao \\
  National Engineering Laboratory for \\Big Data System Computing Technology \\Guangdong Province Key Laboratory \\of Popular High Performance Computers\\ College of Computer Science and \\ Software Engineering \\
  Shenzhen University\\
  Shenzhen 518060, PR China \\
   \And
  Xiao-Min Huang \\
  National Engineering Laboratory for \\Big Data System Computing Technology \\Guangdong Province Key Laboratory \\of Popular High Performance Computers\\ College of Computer Science and \\ Software Engineering \\
  Shenzhen University\\
  Shenzhen 518060, PR China \\
  \And
  Xing-Tong Wu \\
  National Engineering Laboratory for \\Big Data System Computing Technology \\Guangdong Province Key Laboratory \\of Popular High Performance Computers\\ College of Computer Science and \\ Software Engineering \\
  Shenzhen University\\
  Shenzhen 518060, PR China \\
  \And
  Ming-Kai Liu \\
  National Engineering Laboratory for \\Big Data System Computing Technology \\Guangdong Province Key Laboratory \\of Popular High Performance Computers\\ College of Computer Science and \\ Software Engineering \\
  Shenzhen University\\
  Shenzhen 518060, PR China \\
  \And
  Alexandre Vidmer \\
  National Engineering Laboratory for \\Big Data System Computing Technology \\Guangdong Province Key Laboratory \\of Popular High Performance Computers\\ College of Computer Science and \\ Software Engineering \\
  Shenzhen University\\
  Shenzhen 518060, PR China \\
  \And
  Mingyang Zhou \\
  National Engineering Laboratory for \\Big Data System Computing Technology \\Guangdong Province Key Laboratory \\of Popular High Performance Computers\\ College of Computer Science and \\ Software Engineering \\
  Shenzhen University\\
  Shenzhen 518060, PR China \\
  \And
  Yi-Cheng Zhang \\
  National Engineering Laboratory for Big Data System Computing Technology \\Guangdong Province Key Laboratory of Popular High Performance Computers\\ College of Computer Science and Software Engineering \\
  Shenzhen University\\
  Shenzhen 518060, PR China \\
  Department of Physics\\
  University of Fribourg\\
  1700 Fribourg, Switzerland\\
}

\begin{document}
\maketitle

\begin{abstract}
Prediction is one of the major challenges in complex systems. The prediction methods have shown to be effective predictors of the evolution of networks. These methods can help policy makers to solve practical problems successfully and make better strategy for the future. In this work, we focus on exporting countries’ data of the international trading network. A recommendation system is then used to identify the products corresponding to the production capacity of each individual country, but are somehow overlook by the country. Then, we simulate the evolution of the country's fitness if it would have followed the recommendations. The result of this work is the combination combine these two methods to provide insights to countries on how to enhance the diversification of their exported products in a scientific way and improve national competitiveness significantly, especially for developing countries.
\end{abstract}

% keywords can be removed

\section{Introduction}
International trade plays a considerable role in the exchange channels between countries \cite{Almog2015The,Liao2017Ranking}. It is also becoming more important as time goes on. In 1960, international trade accounted for roughly 25\% of a country's total Gross Domestic Product (GDP). Nowadays it accounts for nearly 60\% of countries GDP \cite{WBTrade}.  Historically, various classic economic models were developed to evaluate the dissimilarity of wealth that result from the exportation of diverse goods \cite{Hummels2005The}. Models based on the gravity equation were also developed and showed to be adequate to explain many of the features of the international trade \cite{Fagiolo2010The}. Another approach, The Product Space, attempted to illustrate how the nations will develop in the future by projecting the exports data onto a two-dimensional map and observing the diffusion of the export process \cite{hidalgo2007product}. Economic models mainly consist of commodity profits, geographical relations, comprehensive productivity, economic structure \cite{Tomasello2014The} and a series of macroeconomic elements. In practice, numerous complex external factors e.g. national policies, religious beliefs and the country’s current production capacity, are key components in the international trade. Here, we also recognize that plenty of sociologists have constructed grand theories on the empirical study of Global Economic Development \cite{Portes1993Embeddedness,wallerstein2011modern,sassen2018cities}. These methods and theories were developed by David Snyder, and Edward L. Kick \cite{Snyder1979Structural}, it was the first study of international trade and world economic growth using a network framework in American journal of sociology 1979. In Ref. \cite{kick2011multiple}, the authors emphasized on the importance of the global trade, and the structure of modern world system were addressed by multiple-network analysis. The complex network approach was also used in \cite{Serrano2003Topology}, and it was shown that the international trade network and the World Wide Web both have collaborative characteristics \cite{garcia2014digital}. Indeed, the international trade network is a complex network in terms of structure. With this is in mind, we apply recommendation algorithms that are usually applied on e-commerce systems in order to predict the evolution of the international trade network \cite{lu2011link,LibenNowell2007,vidmer2015prediction}.

Two methods were proposed to assign scores to countries and products using the complex network approach. The basis of these methods is to analyze the relation between countries and their exports, and then to rank the countries according to their economic competitiveness and the exported good with the economic advantage they bring to the exporting countries. The first method, the Method of Reflection (MR), proposed the Economic Complexity Index (ECI) to account for the production characteristics of countries. The countries are ranked according to their exporting capacity. The authors pointed out that the method is able to predict the future economic expansion of countries and that the scores obtained with MR indicate which products should be exported in order to maximize the countries' performance \cite{hidalgo2009building}. In Ref. \cite{caldarelli2012network}, a method based on the Markov chain was proposed. The method showed the need to take into account the nonlinear interactions between exported commodities and national diversity. This lead to the development of the Fitness and Complexity metric \cite{tacchella2012new}. It is an iterative method based on the nonlinearity of the system. This method was shown to be conceptually more grounded than the previous ones and a better predictor of the economic evolution \cite{pietronero2017economic,cristelli2013measuring}. This method was applied to forecast the GDP growth in a recent work. The authors reported that their estimates were on average 25 percent more accurate than were those made by the International Monetary Fund \cite{tacchella2018dynamical}.

In this paper, we start with a detailed definition of the used algorithms, namely the Probabilistic Spreading algorithm \cite{zhou2007bipartite}, the Heat Spreading algorithm \cite{zhou2010solving}, the Degree Increase algorithm \cite{zeng2013trend}, and the Time-aware probabilistic spreading algorithms \cite{vidmer2016role}. We then apply these four algorithms to the international trade network and compare their performance on two different aspects. The first one is the evaluation of the accuracy of the recommendation, which is a standard evaluation of the recommendation algorithms. The second aspect is the evaluation of the algorithms on their ability to recommend products that would improve the countries' fitness. This is obtained by the combination of the recommendation results with the Fitness and Complexity scores to simulate the changes in rankings and fitness values of countries after exporting recommended products. The experimental results confirm the validity of the recommendation algorithm on this task, and show the validity of our approach to tackle this problem.

\section{Materials and Methods}
\subsection{International trade network}
\textbf{2001 to 2015:}
This dataset was cleaned using \textit{harmonization} techniques in Ref.~\cite{gaulier2010baci} and the similar categories were merged together. Additionally, the countries that had no entries recorded for exportation between 2001 and 2015 were removed. After cleaning procedure, the international trade network comprise of 192 countries and 786 commodities. We use the bipartite network approach to represent the data. In this approach, one set of nodes represent the countries, and the second set of nodes represent the commodities. If a country is considered as an exporter of a commodity, a link connects the country's node and the commodity node.

\textbf{RCA:}
The data of the international trade network includes country nodes and commodity nodes. These two types of nodes form a bipartite network. One important aspect of our network representation is that it is binary. Either two nodes are connected, or they are not. We then need a criterion to define if a country can be considered as an exporter of a commodity or not. Indeed, even if countries cannot produce a specific commodity, they might export a very small amount of it. Therefore it should not be considered as an exporter of the commodity. In order to quantify the advantage of a country on a commodity, we use the `Revealed Comparative Advantage' (RCA) as a clear constraint to determine whether a country can be considered as an exporter of certain commodities \cite{balassa1965trade}.
\begin{equation}
RCA_{i\alpha}=\frac{e_{i\alpha}}{\sum_j e_{j\alpha}} \Bigg/ \frac{\sum_\beta
e_{i\beta}}{\sum_{j\beta}e_{j\beta}},
\label{con:rca}
\end{equation}
Where $e_{i\alpha}$ is the total amount of export $\alpha$ for country $i$. If the country $i$ is regarded as an exporter of the commodity $\alpha$, the export amount of the commodity $\alpha$ should occupy a larger proportion in the total amount of the export goods of the country $i$. We set the RCA value to 1 and the country-commodity will have links when this pair of nodes satisfies the condition of $RCA_{i\alpha}\geq1$. These nodes and links together constitute the bipartite network of the international trade activities.

\subsection{Methods}
\subsubsection{Recommendation System}
Recommendation systems aim at recommending the most adequate items for users. In comparison with traditional link prediction, the focus is put on individual nodes rather than individual links. In our case, the main focus of the recommendation process are the countries nodes. For each country node in the system, the algorithms compute a score for every product in the network. If the country is already considered as an exporter of an item, the score corresponding to the product is set to 0 (i.e. a product that is already exported is not recommended). The recommendation list for each country is composed by the top-$L$ products with the highest score for the particular country. We now describe the algorithms used in this work. Note that $L$ is set to 20 in this work and it has been shown to have no impact on the significance of the accuracy \cite{vidmer2015prediction}.

In order to compare the performance of the five recommender algorithms, We choose a year $T$ and predict which additional good the countries exports at year $T+5$ (the testing set) based solely on the data up to year $T$ (the training set).

\textbf{Probabilistic spreading (ProbS):}
For target user $i$, the initial resources are first distributed evenly on the item side and then propagated to the user side through a random walk process \cite{zhou2007bipartite}. In the same way, the resources are then returned to the item side. Both steps are used to allocate resources among neighbors and then spread to the other side. Finally, the score of each item for country $i$ is obtained. The initial resource vector is represented as $f^{(i)}$ and its elements as $f_\alpha^{(i)}$=$a_{i\alpha}$. The final resource values $s_\alpha^{(i)}$ can be written as
\begin{equation}
  s_\alpha^{(i)}=\sum\nolimits_{\beta =1}^IW_{\alpha \beta }f_\beta ^{(i)}
 \end{equation}
The elements of the redistribution matrix $W$ are derived from probabilistic spreading process.
\begin{equation}
W_{\alpha\beta}=\frac{1}{k_\beta }\sum\nolimits_{j =1}^U\frac{a_{j\alpha }a _{j\beta }}{k_j}
 \end{equation}
where $I$ denotes the number of products and $U$ the number of countries. $a_{j\alpha}a _{j\beta }$ represents path from item $\beta$ to item $\alpha$ through country $j$ in two random walks. The ProbS algorithm employs a column-normalized transition matrix. The partition of $k_\beta$ and $k_j$ corresponds to the uniform partition of resources between all links from nodes $\beta$ and $j$ \cite{yu2016network}.

\textbf{Heat spreading (HeatS):}
The Heats spreading algorithm evolved from the Probabilistic spreading algorithm. These two methods are similar in structure and both use random walk processes to the redistribute initial resources. Compared with the ProbS algorithm, resources spread more evenly in the Heats spreading method and items with only few connections usually benefit from a higher score than with the ProbS algorithm \cite{zhou2010solving}.

The initial resource vector is set to $f^{(i)}$ according to user's item collection, the elements $f_\alpha^{(i)}$=$a_{i\alpha}$ can be regarded as the temperature value of the item. Unlike the ProbS algorithm, which uses column normalization transformation matrix, the HeatS algorithm uses row normalization. The spreading process is represented by a matrix as follows, where $ W{}'=W^T$.
\begin{equation}
 W{}'_{\alpha\beta}=\frac{1}{k_\alpha}\sum\nolimits_{j =1}^U\frac{a_{j\alpha }a _{j\beta }}{k_j}
 \end{equation}
Resource received by the user $i$ is equal to average resource owned by the user's collected item. Similarly, item side receives resources transmitted from user through the averaging process. The item’s score is calculated as:
\begin{equation}
  h_\alpha^{(i)}=\sum\nolimits_{\beta =1}^IW{}'_{\alpha \beta }f_\beta ^{(i)}
 \end{equation}

\textbf{Degree Increase (DI):}
The time information is often overlooked in the evolution of complex networks. In fact, time plays a crucial role in the evolution of information networks \cite{Crane2008Robust,szabo2010predicting,ren2016characterizing}. The combination of recommendation systems with time dynamics improve the recommendation and allows to perform better predictions. It has been proven that Degree Increase (DI) method can accurately predict the prevalence of items in the future. The increase in popularity of item $\alpha$ within time window $\tau$ is:
\begin{equation}
\Delta k_\alpha (t,\tau )=k_\alpha (t)- k_\alpha (t-\tau )
\end{equation}
where item degree $k_\alpha \left ( t \right )=\sum\nolimits_iA_{i\alpha}\left ( t \right )$ corresponds to the number of users who have collected item $\alpha$. The final item score is expressed as:
\begin{equation}
\Delta{k{}'}_\alpha (t,\tau )=\Delta k_\alpha (t,\tau )+\varepsilon k_\alpha (t)
\end{equation}
In practice, the value of $\varepsilon$ must be small enough to ensure that the ranking of commodity popularity growth $\Delta k_\alpha (t,\tau )$ does not change.

\textbf{Time-aware probabilistic spreading (TProbS):}
The ProbS method is characterized by the fact that spread of resources is cumulative, that is, the more popular items are more likely to get high scores and recommend to users. The Time-aware probabilistic spreading method integrate the ProbS method with the DI method. It writes mathematically as:
\begin{equation}
u_\alpha^{(i)} =s_\alpha^{(i)}\left(\frac{\Delta{k{}'}_\alpha (t,\tau )}{k_\alpha (t)}\right)^\theta
\end{equation}
where the parameter $\theta$ is an additional parameter to define the length of past time window.

\subsubsection{Fitness and Complexity metrics}
The Fitness and Complexity metrics are used to measure the competitiveness of countries and the quality of exported products. The algorithm consists of two nonlinear coupling equations \cite{tacchella2012new,LiaoH2018AComparative}, which eventually reach a fixed value through iterative methods and equations is defined as
\begin{equation}
F_i^{n}=\sum_{\alpha=1}^{I} a_{i\alpha} Q_\alpha^{n-1}
\label{eq:Fit}
\end{equation}
\begin{equation}
Q_\alpha^{n} =\frac{1}{\sum_{i=1}^{U} a_{i\alpha} 1/F_i^{n-1}}
\label{eq:Comp}
\end{equation}
where $F_i^n$ indicates the fitness of country $i$ after $n$ iterations. In Ref. \cite{pugliese2016convergence}, the convergence of the algorithm and its stopping condition were demonstrated. The higher the fitness value, the more advantageous the variety and complexity of the goods exported by the country. The study \cite{pugliese2017complex} shows, the weak economies can analyze how to get out of the poverty trap and increase the diversification of their exports via fitness metrics. Complexity $Q_\alpha^n$ cannot simply calculate from average of countries' fitness. Successful countries export almost all products and it is unable to infer the complexity of each product from their export data (for example, our data shows, a total of 786 products were included in the International Trade Network from 2001 to 2015, and the United States exported a total of 775 kinds of products). Therefore, the complexity of product should be measured in a non-linear way, namely, it is essential to reduce contribution of successful countries. Assuming that product $\alpha$ possess two exporter $i$ and $j$ with fitness values of $0.2$ and $15$, respectively, the complexity of the product would be $0.197$. If the two exporter $i$ and $j$ with fitness values of $10$ and $15$, the complexity would be $6$. This two examples verify the fact value of complexity depends mainly on the worst exporter.

\subsubsection{Precision and Recall metrics}
In general, recommendation algorithms are compared in their ability to predict the future. In order to evaluate their accuracy, we use two metrics, namely the Precision, and the Recall metrics. We are not interested in the accuracy of the algorithms per se, but the accuracy of the recommendation results is an important evaluation indicator. A higher accuracy indicates that the exporting countries possess the capacity to produce the recommending commodities. Indeed, the countries should be able to produce the recommended commodities in the near future, otherwise the recommendation process would be meaningless.
\\
\textbf{Precision:} For each country, the individual precision is measured as the fraction of recommended products that are eventually exported. If $n_i$ is the number of products eventually exported by country $i$ and that are actually recommended, then the precision for country $i$ is $P_i=n_i/L$. The Precision $P$ is the average of $P_i$ over every countries.
\textbf{Recall:} Recall is similar to precision, but we use the number of newly exported goods instead of the fixed recommendation list $L$. If country $i$ exports $E_i$ additional items in the testing set, the individual recall reads $R_i=n_i/E_i$. The Recall is the average of $R_i$ over every countries.

\subsection{Addition of products to countries' export basket}
After having obtained the recommendation list from the previously described recommendation methods, we add the goods that are at the top of the recommendation lists to the respective countries' exports basket. Then, the fitness value of each country is reevaluated and its change is recorded. However, the calculations of fitness is an iterative nonlinear process, which is coupled with the complexity of the products. Thus, in order to evaluate the changes brought by adding the products to the country's export basket, only one country is modified at a time. In other words, for a given country $i$, only its recommendation list is added to its exports basket and the rest of the network is left untouched. For instance, if we have a high complexity product to the export basket of a low fitness country, the complexity of the product might be greatly reduced, which will have a strong impact on the rest of the fitness and complexity values. The simulation process is described as below. For each country, we add the $L$ products in its recommendation list to its export basket. For each product $\alpha$ in the top $L$ part of the recommendation list of country $i$, we add a link between the country $i$ and the product $\alpha$ in the data. The export volume of product $\alpha$ from country $i$ must satisfy the condition of $RCA\geq1$.

\section{Results}
\subsection{Accuracy of the recommendation algorithms}
\begin{figure}[!htb]
\centering
\includegraphics[width=0.9\columnwidth]{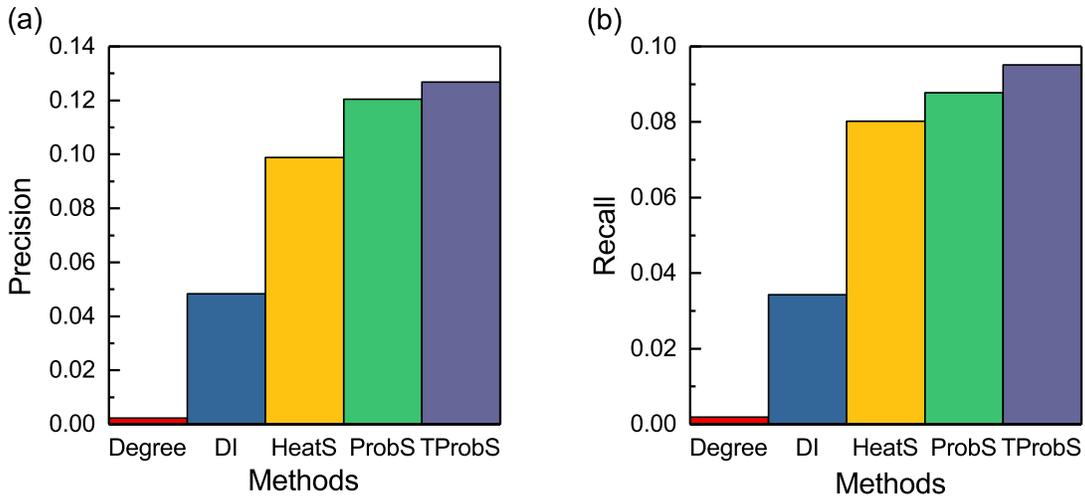}
\caption{A comparison of Recall and Precision for the five algorithms. The recommendation is performed at year $T$ for year $T+5$, with $T$ ranging from year 2001 to 2010. The results shown are averaged over this time period. For TProbS, we set a value of $\theta=0.2$. The Degree method simply ranks the products according to their degree. All subsequent experiments were based on this recommendation.}
\label{fig:pr}
\end{figure}
The first step in the comparison of the algorithms is to compare their accuracy. We perform the predictions from years 2001 to 2010. The results are shown on \prettyref{fig:pr}.
First, we see that the top performing method is the TProbS method. This is not surprising as this is the one including both the strength of ProbS and time information. The parameter $\theta=0.2$ is constantly the one giving the highest recall in our simulation, and thus is fixed to this value. Obviously, this lessen the impact of degree increase on the network, which is compatible with the expected behavior of countries development. They should not all focus on the same products, but they still follow the ongoing trends. As noted in \cite{vidmer2015prediction}, the HeatS method performs surprisingly well. This method usually performs poorly in online e-commerce network \cite{Zhouothers2010}, but has the advantage of recommending products with lower degree. DI and Degree both perform poorly, which is expected as they do not take into account the production capabilities of countries.

\subsection{Tier division}
\begin{figure}[ht]
\centering
\includegraphics[width=0.9\columnwidth]{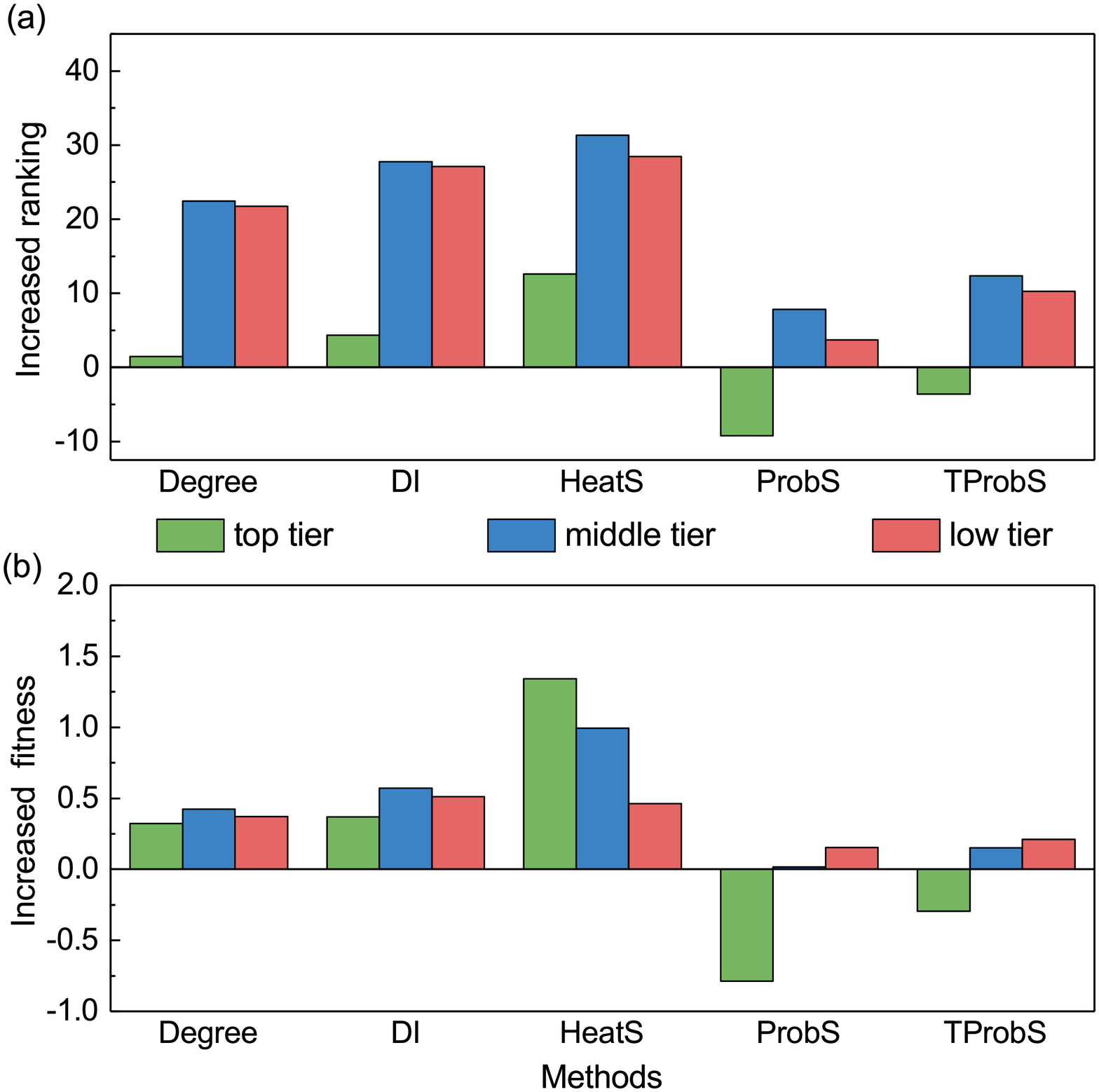}
\caption{
Panel (a) is the average increase of fitness ranking for the three different tiers of countries for the time period 2008-2015. The number of goods added for each country is set to $L=20$. Panel (b) is the average increased fitness value of those three tiers.}
\label{fig:overall-rank}
\end{figure}

An interesting study is to investigate the impact of the methods on countries with different fitness rank. It is not certain that a country that belongs to the
top of the fitness list will behave the same than a country that belongs at the bottom of the list. In order to study this effect, we divide the countries into three different categories according to their fitness rank.
The three categories hold the same number of countries and thus are denoted as top tier, middle tier, and low tier. For each category, we compute the average increased ranking and increased fitness after the addition of the recommended exports. The results are shown in \prettyref{fig:overall-rank}. The countries that benefits the most from these recommendation are middle and low tier countries. In panel (a), the most interesting result is in HeatS. The increased ranking by following its recommendation are much higher than those of TProbS. This is especially true with top tier countries, as following the recommendation of TProbS would even lower a country's fitness. This comes from the fact that top tier countries need to innovate and produce goods for which there are only a few competitors. While for middle and low tier countries, it is sufficient to produce additional items. Note that we try adding time to HeatS: the recall improved to the level of TProbS, but the improvement of fitness was lower than that of HeatS. So we decided to keep only HeatS and TProbS for simplicity.

From panel (b), we see that the fitness of top tier countries is the one improving the most, which shows that the recommendation algorithm indeed recommend different complex products in line with the ability of the country. For top tier countries, the difference of fitness between two countries is large. While for lower tier countries, the fitness difference between two countries is much smaller. This fact explains the result that top tier countries' fitness improvement is significant but their ranking is only slightly improved, while the low tier countries can increase the ranking a lot in the case of less improvement in fitness.
To illustrate better our recommendation results, we select three countries in each tier. Due to the limit of the page size, we only choose the five products with the highest score according to TProbS and HeatS for each country. The products recommended by TProbS and HeatS are shown in \prettyref{tab:tprobs} and \prettyref{tab:heats}, respectively. Both tables show the products with different complexity for various tier countries.

\subsection{Evolution of countries' fitness}
\begin{figure}[ht]
\centering{}
\includegraphics[width=0.95\columnwidth]{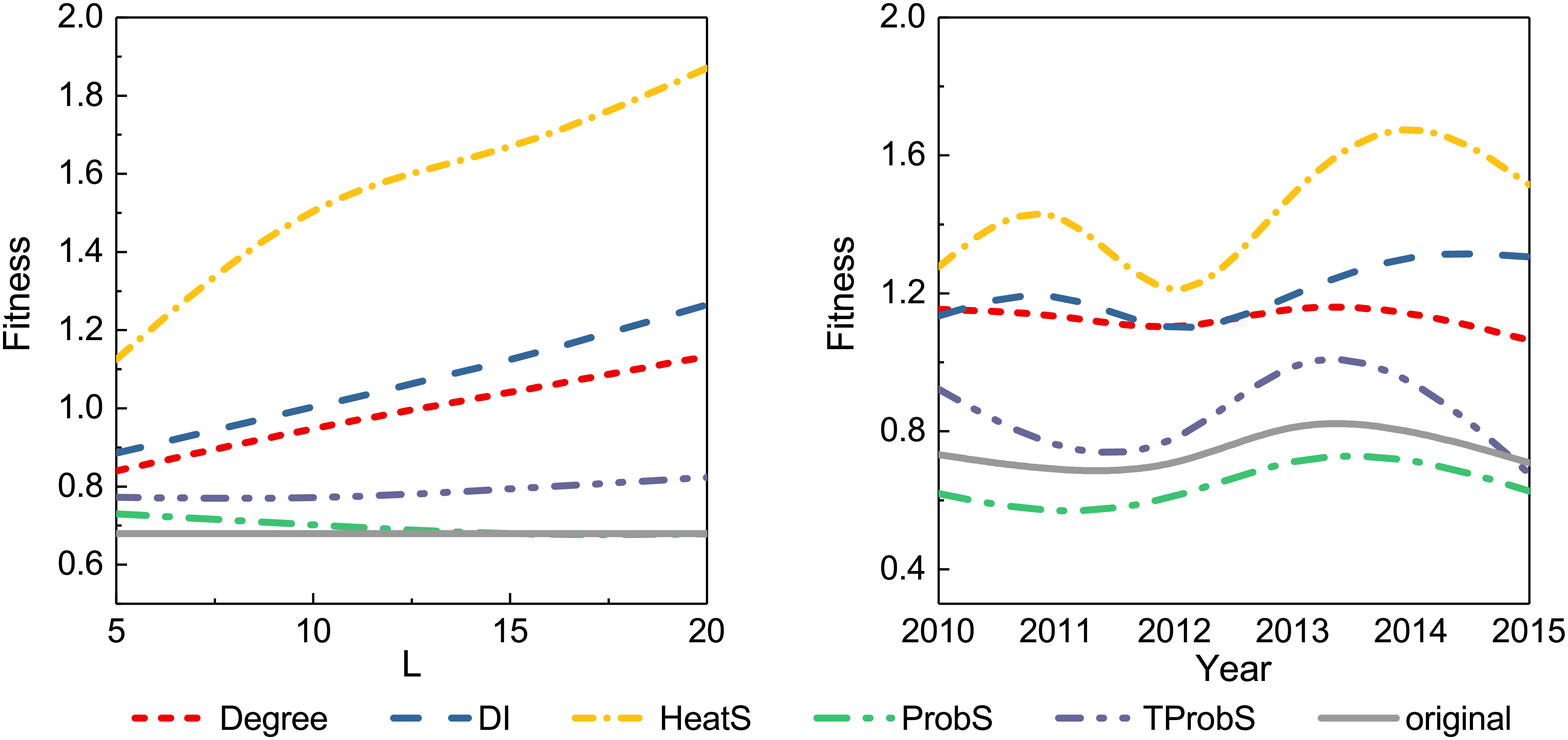}
\caption{Panel (a) shows that fitness values as a function of the length of the recommendation list $L$. The results are average over 30 randomly selected countries and over the 15 years spanned by the data. Time evolution of fitness for the thirty selected middle tier countries that are shown in panel (b). The length of the recommendation list is set to $L=20$.
}
\label{fig:average30}
\end{figure}

In order to directly study the effect of each algorithm on the country’s fitness value, we randomly select 30 countries in the middle tier and compare the evolution in the average fitness of these countries for the different methods. We compare the normal evolution of countries and the one recommended by these algorithms. The results are shown in \prettyref{fig:average30} (panel (a)). For any length of the recommendation list $L$, HeatS is the top performing by a great margin compared to other methods. For all methods, ranking are consistent for every choice of $L$, and so the choice of the value is not meaningful, as long as it is reasonably small compared to the total number of items. By looking at the evolution with the length of the recommendation list $L$, we find that the results are consistent and if the country should follow the recommendation, even if it can add only few products to its export basket. Following the recommendation made by HeatS seems to be adequate in terms of technological requirements (i.e. the countries have the required technology) as well as the best path to increase the country's fitness.
In \prettyref{fig:average30} (panel (b)), the results are shown for different years and for fixed $L$. Though some methods are better at specific year than other, it is clear that HeatS is always the top performing method, followed by Degree and DI and then TprobS. This is good news as it shows the robustness of our method, both in the isolated case and in the real evolution of countries' exports.

\begin{table}[htp]
%~ \centering
\footnotesize
\begin{center}
\begin{tabular}{lll}
\hline
Tier & Country & Recommended products (top 5) \\
\hline
& Vanuatu & Cigarettes \\
&&Fruit, fresh or dried\\
&&Plants and parts of trees used in perfumery; in pharmacy; etc\\
&&Sawlogs and veneer logs, of non-coniferous species\\
&&Non-alcoholic beverages\\
\cline{2-3}
Low tier& Tonga & Spices, except pepper and pimento\\
&& Sugar confectionery and preparations, non-chocolate\\
&& Fish, dried, salted or in brine; smoked fish\\
&& Cement\\
&& Wood, simply shaped\\
\cline{2-3}
& Dominica & Meal and flour of wheat and flour of meslin\\
&& Cask, drums, etc, of iron, steel, aluminium, for packing goods\\
&& Packing containers, box files, etc, of paper, used in offices\\
&& Beer made from malt\\
&& Plastic packing containers, lids, stoppers and other closures\\
\hline
& Norway & Animals, live (including zoo animals, pets, insects, etc)\\
&&Fuel wood and wood charcoal\\
&&Bovine and equine hides, raw, whether or not split\\
&&Bones, ivory, horns, coral, shells and similar products\\
&&Soaps, organic products and preparations for use as soap\\
\cline{2-3}
Middle tier& Mauritius &Skirts\\
&& Vegetable products roots and tubers, fresh, dried\\
&& Leather of other hides or skins\\
&& Footwear\\
&& Other outer garments\\
\cline{2-3}
& Kyrgyzstan & Oil seeds and oleaginous fruits\\
&& Fish, fresh or chilled, excluding fillet\\
&& Eggs, birds', and egg yolks, fresh, dried or preserved, in shell\\
&& Fruit or vegetable juices\\
&& Jams, jellies, marmalades, etc, as cooked preparations\\
\hline
& Germany & Iron, steel, aluminium reservoirs, tanks, etc, capacity 300 lt plus\\
&&Non-domestic refrigerators and refrigerating equipment, parts\\
&&Insulated electric wire, cable, bars, etc\\
&&Bottles etc of glass\\
&&Railway or tramway sleepers\\
\cline{2-3}
Top tier& Italy& Refined sugar etc\\
&& Fuel wood and wood charcoal\\
&& Sugar confectionery and preparations\\
&& Manufactures of mineral materials (other than ceramic)\\
&& cotton, not elastic nor rubberized\\
\cline{2-3}
& Switzerland & Structures and parts of, of iron, steel; plates, rods, and the like\\
&& Manufactures of mineral materials (other than ceramic)\\
&& Other furniture and parts thereof\\
&& Organic surface-active agents\\
&& Miscellaneous articles of base metal\\
\hline
\end{tabular}
\end{center}
\caption{The top five recommended products by TProbS algorithms for three randomly selected countries in different tiers.}
\label{tab:tprobs}
\end{table}

\begin{table}[htp]
%~ \centering
\footnotesize
\begin{center}
\begin{tabular}{lll}
\hline
Tier & Country & Recommended products (top 5) \\
\hline
& Vanuatu & Sugars, beet and cane, raw, solid \\
&&Natural rubber latex; natural rubber and gums\\
&&Jute, other textile bast fibres, raw, processed but not spun\\
&&Ores and concentrates of uranium and thorium\\
&&Nuts edible, fresh or dried\\
\cline{2-3}
Low tier& Tonga & Manila hemp, raw or processed but not spun\\
&& Coconut (copra) oil\\
&& Cocoa beans, raw, roasted\\
&& Waxes of animal or vegetable origin\\
&& Palm nuts and kernels\\
\cline{2-3}
& Dominica & Potassium salts, natural, crude\\
&& Distilled alcoholic beverages\\
&& Surveying, navigational, compasses, etc, instruments, nonelectrical\\
&& Figs, fresh or dried\\
&& Beer made from malt\\
\hline
& Norway & Crustaceans and molluscs, fresh, chilled, frozen, salted\\
&&Asbestos\\
&&Radio-active chemical elements, isotopes etc\\
&&Ships, boats and other vessels\\
&&Iron ore agglomerates\\
\cline{2-3}
Middle tier& Mauritius &Fabrics, woven\\
&& Fish, dried, salted or in brine; smoked fish\\
&& Articles of apparel, clothing accessories of plastic or rubber\\
&& Household appliances, decorative article, etc, of base metal\\
&& Headgear and fitting thereof\\
\cline{2-3}
& Kyrgyzstan & Meat of sheep and goats, fresh, chilled or frozen\\
&& Iron ore and concentrates, not agglomerated\\
&& Coffee green, roasted; coffee substitutes containing coffee\\
&& Cut flowers and foliage\\
&& Railway or tramway sleepers (ties) of wood\\
\hline
& Germany & Photographic and cinematographic apparatus and equipment\\
&&Organo-sulphur compounds\\
&&Parts of the pumps and compressor\\
&&Orthopaedic appliances, hearing aids, artificial parts of the body\\
&&Phenols and phenol-alcohols, and their derivatives\\
\cline{2-3}
Top tier& Italy& Digital central storage units, separately consigned\\
&& Children's toys, indoor games, etc\\
&& Optical instruments and apparatus\\
&& Fabrics of glass fibre (including narrow, pile fabrics, lace, etc)\\
&& Printing presses\\
\cline{2-3}
& Switzerland & Electro-medical equipment\\
&& Chemical products and preparations\\
&& Parts of steam power units\\
&& Spectacles and spectacle frames\\
&& Other chemical derivatives of cellulose; vulcanized fibre\\
\hline
\end{tabular}
\end{center}
\caption{The top five recommended products by HeatS algorithms for three randomly selected countries in different tiers.}
\label{tab:heats}
\end{table}

\begin{figure}[h]
\centering
\includegraphics[width=0.9\columnwidth]{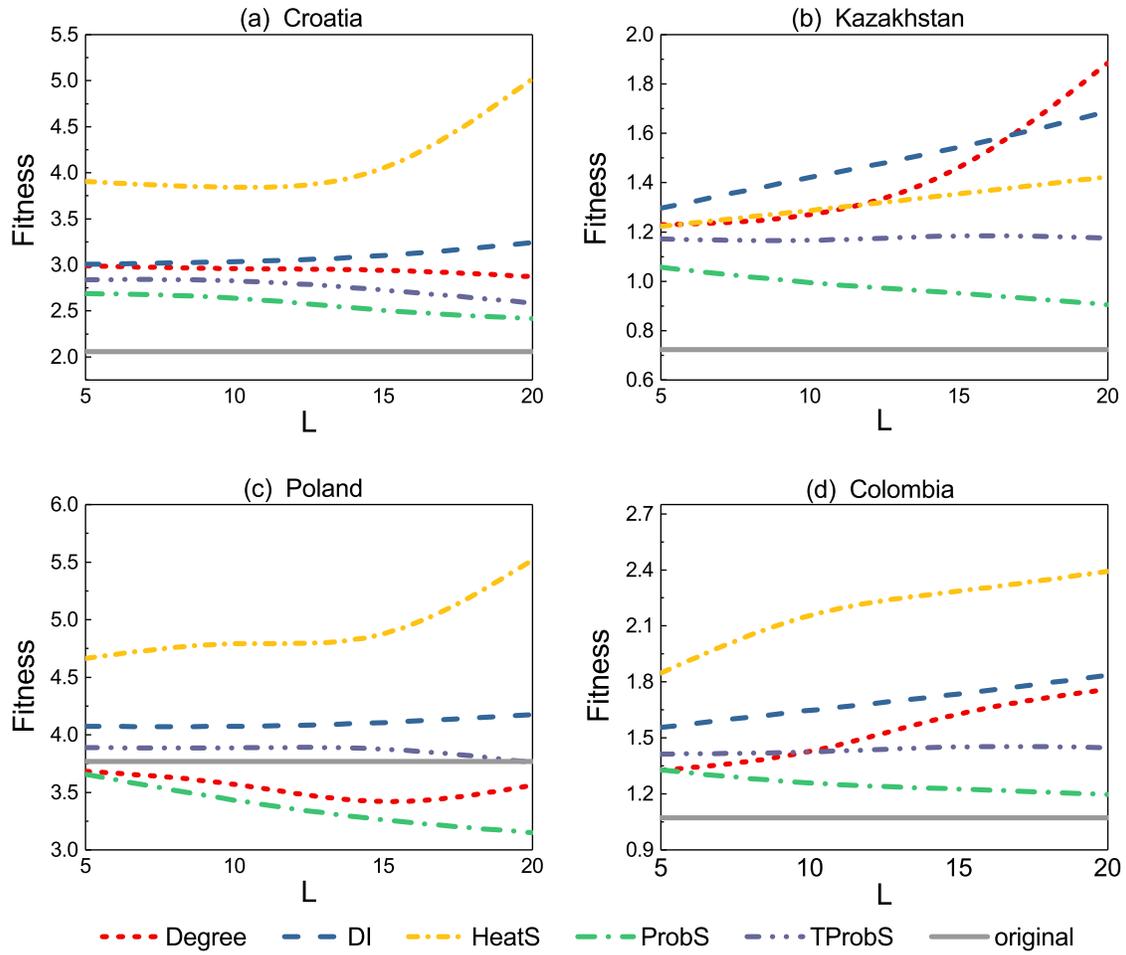}
\caption{Detailed view of the fitness as a function of the length of the recommendation list $L$ for four randomly selected countries. The original fitness values of four countries Croatia, Kazakhstan, Poland and Colombia are 2.0563, 0.7234, 3.7678 and 1.0714 respectively.
}
\label{fig:fit17}
\end{figure}

\begin{figure}[h]
\centering
\includegraphics[width=0.9\columnwidth]{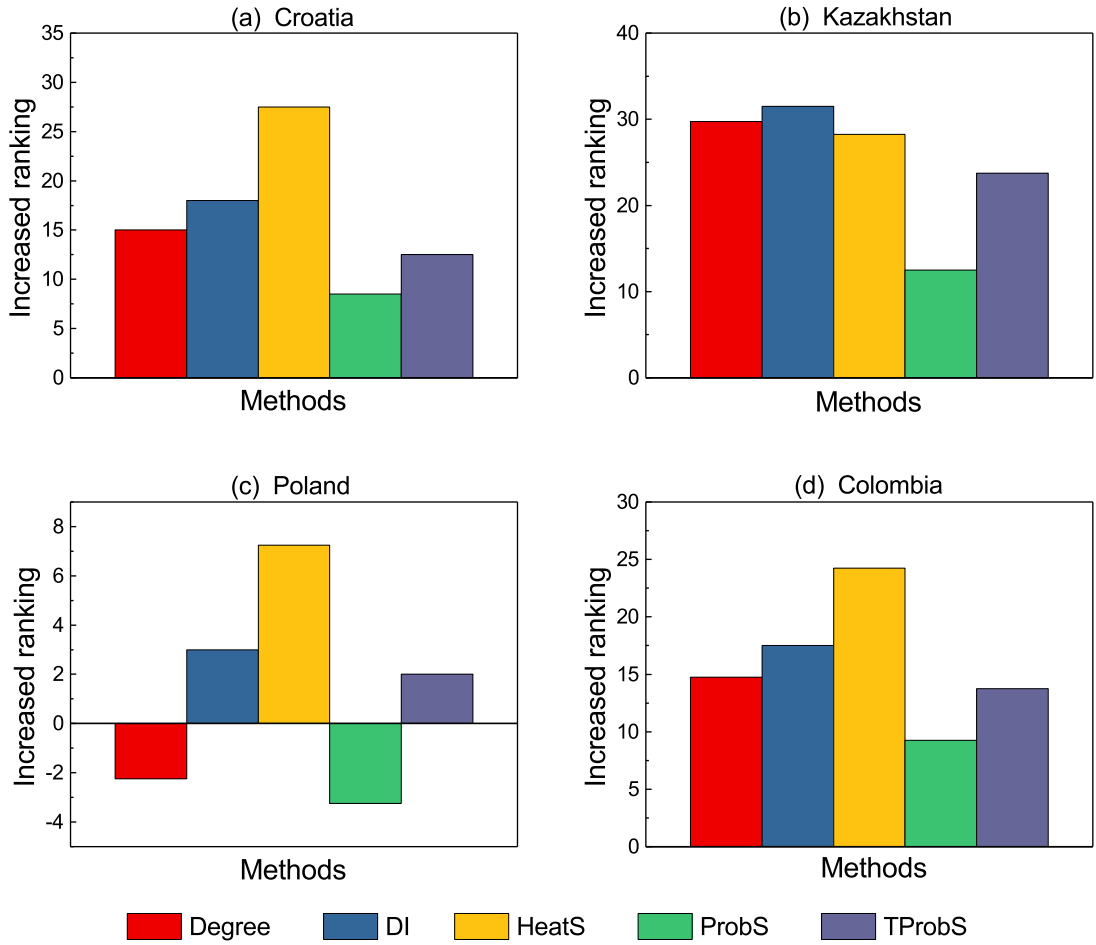}
\caption{Comparison of the increased ranking of the four selected countries. The increased ranking of each method is averaged over different lengths of the recommendation list $L$.}
\label{fig:rank}
\end{figure}

We are also interested in the individual behavior of the fitness evolution. We randomly select four countries, and study the effect of the method on the evolution of the fitness values. The comparison of the five methods as a function of $L$ and the original fitness is shown in \prettyref{fig:fit17}. Again, HeatS is the top performing method, except for Kazakhstan. For this country, DI and Degree are the top performing methods due the low Fitness of this country. The original fitness values corresponding to the four countries Croatia, Kazakhstan, Poland and Colombia are 2.0563, 0.7234, 3.7678 and 1.0714 respectively. While we saw in \prettyref{fig:average30} that HeatS is always best on average, it's different when considering individual countries with especially low production. The impact of the recommendation on the increased ranking of the four countries is shown in \prettyref{fig:rank}. We see in \prettyref{fig:rank} (panel (b)) that even if HeatS is not the best method, the difference in ranking is quite small, while in other panels it clearly outperforms other methods.

\subsection{Real production ability}

\begin{figure}[h]
\centering
\includegraphics[width=0.9\columnwidth]{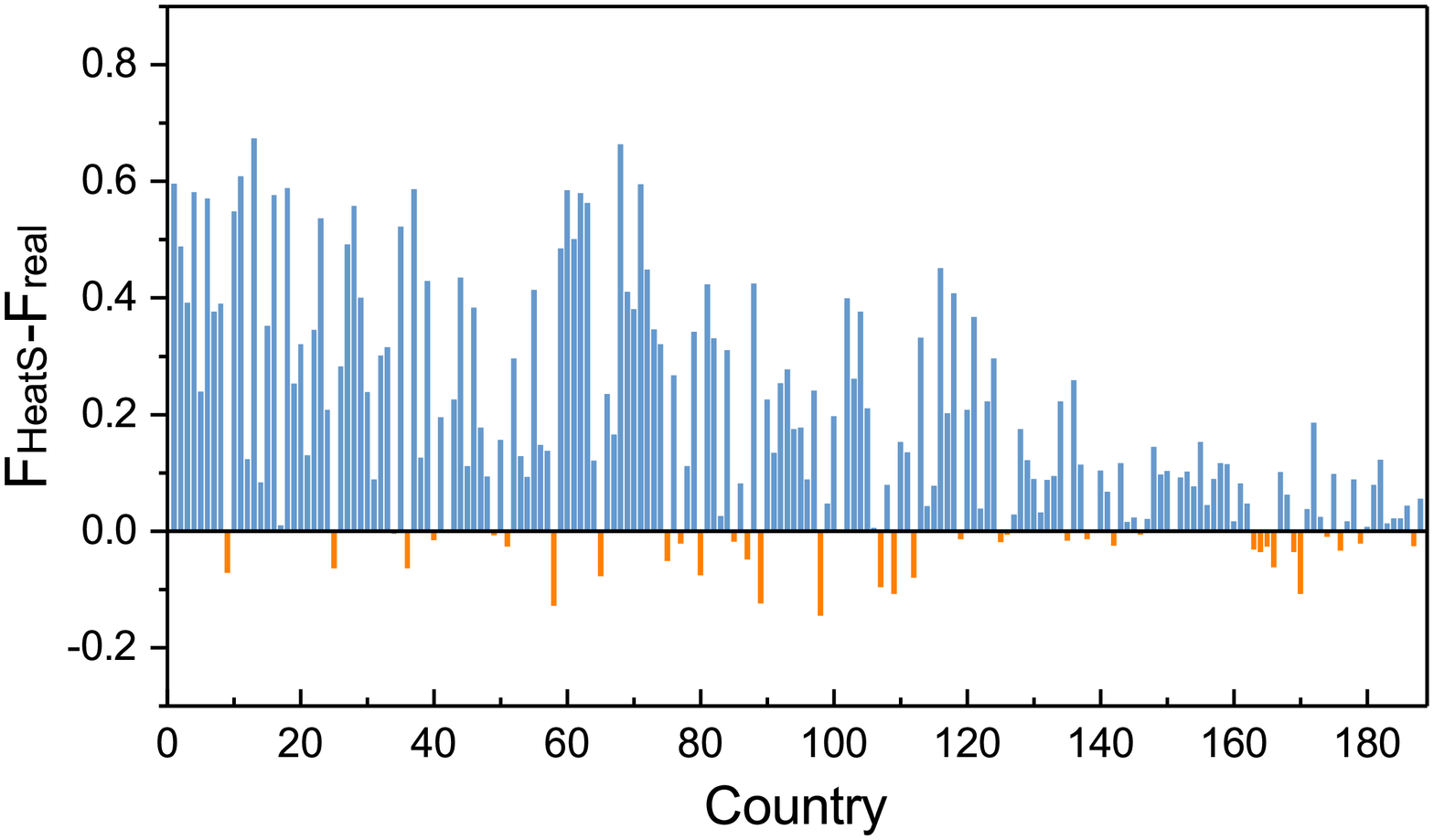}
\caption{ Difference of fitness resulting from following recommendation of HeatS and real evolution. The countries are sorted according to their fitness value, country 0 being the one with highest average fitness. Each bar represents a country. The recommendation is performed at year $T$ for year $T+5$, with $T$ ranging from year 2001 to 2010. 143 countries of the 181 countries would see their fitness improve by following the recommendation of HeatS.}
\label{fig:heats}
\end{figure}

\begin{figure}[!h]
\centering
\includegraphics[width=0.9\columnwidth]{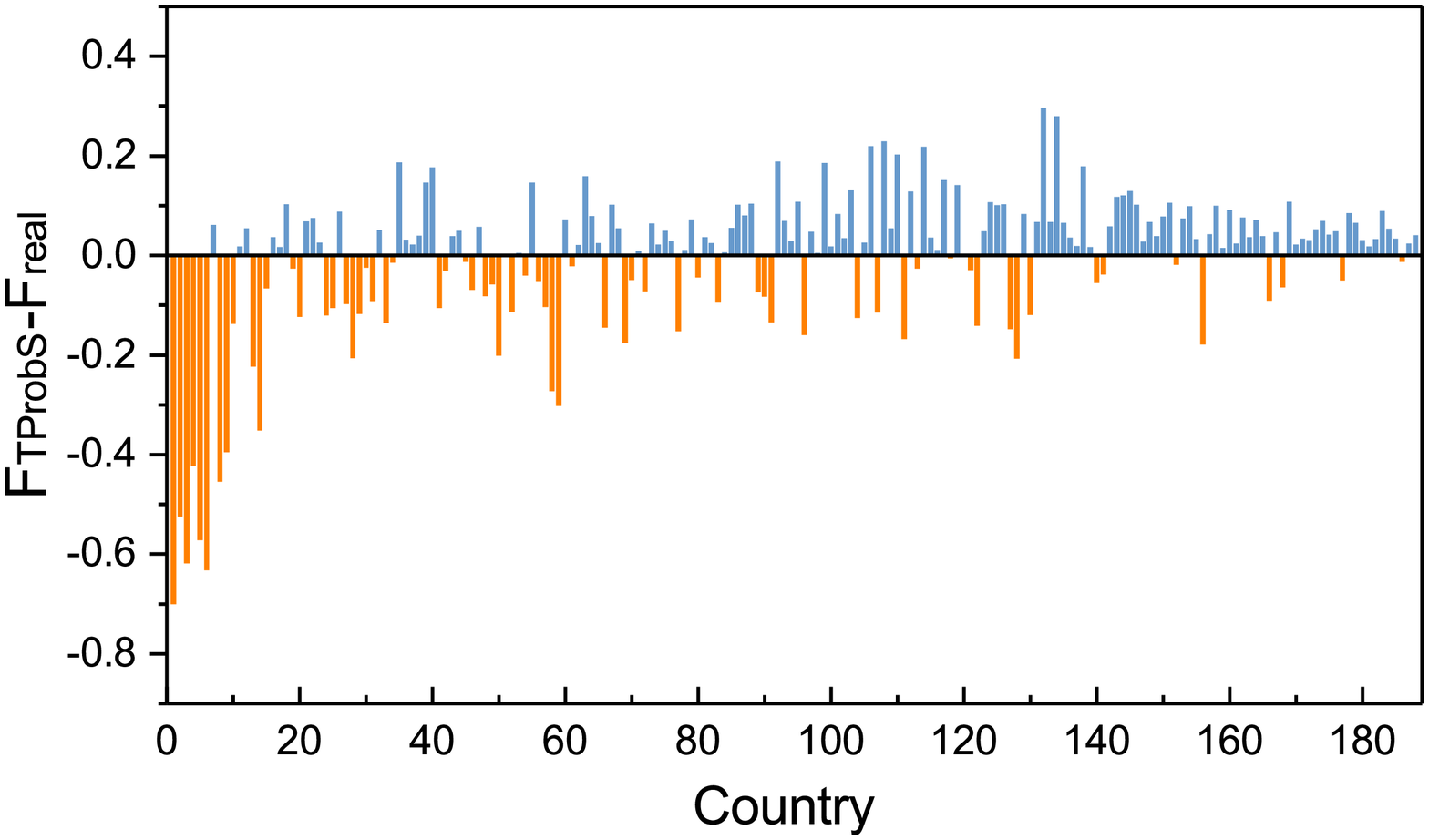}
\caption{Difference of fitness resulting from following recommendation of TProbS and real evolution. The countries are sorted according to their fitness value, country 0 being the one with highest average fitness. Each bar represents a country. The recommendation is performed at year $T$ for year $T+5$, with $T$ ranging from year 2001 to 2010. 120 countries of the 181 countries would see their fitness improve by following the recommendation of TProbS.}
\label{fig:tprobs}
\end{figure}

In the previous results, we fixed a parameter $L$ and assigned the same number of new products to every countries. However, in reality some countries produce many new products at a specific year, while some others struggle more. To reflect this, we use a dynamic length of the recommendation $L_i$, where $L_i$ is the length of the recommendation list for country $i$, equal to its number of new products between time $T+5$ and $T$. We build a `virtual network' corresponding exactly to the real one at $T+5$, except for one country $i$, for which we replace the links that appears between $T$ and $T+5$ by those of the recommendation list. This ensures that the network is of constant size. The results for HeatS are shown in \prettyref{fig:heats}. We see that HeatS improves greatly the fitness of most countries that follow its recommendation. Only a few countries see their fitness decreasing compared to their original evolution. As a comparison, we see in \prettyref{fig:tprobs} that TProbS is of no use for top tier countries, but for middle and especially low tier countries they might benefit from it. While lower than for HeatS, the recommendation of TProbS are made of more widespread technologies and so might be easier to follow for the low fitness countries.
\section{Conclusion}
In this study, we extended the traditional recommendation, which are usually applied on social networks, to the international trade. Thanks to recent advances in measuring the potential countries' evolution based solely on the network structure, we designed a method which aims to help countries to find a suitable evolution path among all the possible ones. The study suffers two main limitations. The first one is that the exports categories are roughly defined.  Only about 786 categories are present in the dataset. This leads to some categories containing very varied products, such as \textit{Iron based goods}. The second one is that there are important restrictions, or conversely support,  to the trade between countries. For instance, USA, Canada, and Mexico recently signed an agreement to open the market of Dairy and Cars parts. On the opposite, countries might limit their sensitive exports, such as military goods, to specific countries. While the first limitation is difficult to address due to the data limitations, the second one can be added by weighting differently the relationships between countries on specific products.

At the same time, the recommendation proved to grasp the countries technological evolution by being able to correctly predict the future, and the method of Fitness and Complexity has been shown in \cite{Tacchella2013Economic,cimini2014scientific,mariani2015measuring} to uncover hidden features of the countries' evolution. It is important to note that recommendation method should agree with the similar capabilities of a country \cite{pugliese2017unfolding}. Those two ingredients together mixed with our results show remarkable evidence that our methods as supplementary message could help to design objective metrics in order to facilitate the work of policy makers and encourage the development of technology towards the better economic goal.

\section*{Acknowledgements}
We acknowledge financial support from the National Natural Science Foundation of China grant number 61803266, 11547040 and 61703281; the Guangdong Province Natural Science Foundation of China grant number 2016A030310051 and 2017A030310374; Foundation for Distinguished Young Talents in Higher Education of Guangdong grant number 2015KONCX143; the Shenzhen Fundamental Research Foundation grant number JCYJ20150529164656096 and JCYJ20170302153955969; the Natural Science Foundation of SZU grant number 2016-24; Guangdong Pre-national project grant number 2014GKXM054.
\newpage
\pagebreak
\newpage

\pagebreak
\newpage

\pagebreak
\newpage

\bibliographystyle{plain}

\end{document}